\documentclass[preprint,12pt]{elsarticle}

\usepackage{amssymb}
\journal{Pervasive and Mobile Computing}

\begin{document}

\clubpenalty=300
\widowpenalty=300

\newcommand{\al}[1]{\footnote{\textbf{Antonio: #1}}}
\newcommand{\mm}[1]{\footnote{\textbf{Mirco: #1}}}
\newcommand{\ma}[1]{\footnote{\textbf{Manlio: #1}}}

\begin{frontmatter}

%% Title, authors and addresses

%% use the tnoteref command within \title for footnotes;
%% use the tnotetext command for the associated footnote;
%% use the fnref command within \author or \address for footnotes;
%% use the fntext command for the associated footnote;
%% use the corref command within \author for corresponding author footnotes;
%% use the cortext command for the associated footnote;
%% use the ead command for the email address,
%% and the form \ead[url] for the home page:
%%
%% \title{Title\tnoteref{label1}}
%% \tnotetext[label1]{}
%% \author{Name\corref{cor1}\fnref{label2}}
%% \ead{email address}
%% \ead[url]{home page}
%% \fntext[label2]{}
%% \cortext[cor1]{}
%% \address{Address\fnref{label3}}
%% \fntext[label3]{}

\title{Interdependence and Predictability\\of Human Mobility and Social Interactions}

%% use optional labels to link authors explicitly to addresses:
%% \author[label1,label2]{<author name>}
%% \address[label1]{<address>}
%% \address[label2]{<address>}

\author{Manlio De Domenico}\ead{m.dedomenico@cs.bham.ac.uk}
\author{Antonio Lima}\ead{a.lima@cs.bham.ac.uk}
\author{Mirco Musolesi}\ead{m.musolesi@cs.bham.ac.uk}

\address{School of Computer Science, University of Birmingham\\Edgbaston B15 2TT, Birmingham, United Kingdom}

\begin{abstract}

% MAX 100 words!!!

Previous studies have shown that human movement is predictable to a certain extent at
different geographic scales. Existing prediction techniques exploit only the past history of the person taken into consideration as input of the predictors.

In this paper, we show that by means of multivariate nonlinear time series prediction techniques it is possible to increase the forecasting accuracy by considering movements of friends, people, or more in general entities,
with correlated mobility patterns (i.e., characterised by high mutual information) as inputs. Finally, we evaluate the proposed techniques on the Nokia Mobile Data Challenge and Cabspotting datasets.

%that using pairs of people with correlated mobility improves the forecasting
%accuracy, when using multivariate nonlinear predictors. 

\end{abstract}

\begin{keyword}
%% keywords here, in the form: keyword \sep keyword
mobility prediction \sep mutual information \sep nonlinear timeseries analysis

%% MSC codes here, in the form: \MSC code \sep code
%% or \MSC[2008] code \sep code (2000 is the default)

\end{keyword}

\end{frontmatter}

%%
%% Start line numbering here if you want
%%
% \linenumbers

%% main text
\section{Introduction}
\label{sec:introduction}

The study of the interdependence of human movement and social ties of
individuals is one of the most interesting research areas in computational
social science~\cite{computationalSocialScience09}.  Previous studies have
shown that human movement is predictable to a certain extent at different
geographic scales~\cite{BHG:Scaling,SMML11:nextplace,Song:Limits}. The potential
uses of these prediction techniques are various, including practical ones, such as content
dissemination of location-aware information, e.g., targeted advertisements in sponsored
mobile applications or in search results performed from mobile
phones~\cite{LGKO04:Bluetooth}.

In this paper we show how it is possible to improve mobility prediction by
exploiting the correlation between movements of individuals. It is possible to exploit such
correlations for prediction and inference of aspects related to user behaviour,
namely their movements and their social interactions (either physical and
distant). In particular, in our analysis we exploit and adapt the concept of
\textit{mutual information}~\cite{infotheory} in order to quantify correlation
and provide a \textit{practical} method for the selection of additional data to
improve the accuracy of movement forecasting. We also show how this quantity
correlates to different types of social interactions of friends and
acquaintances. This paper extends the findings presented in our submission~\cite{DMM12:Interdependence} to the Nokia Mobile Data Challenge competition~\cite{NokiaMDC}.

More specifically, the contributions of this work can be summarised as follows:

\begin{itemize}
\item We first show that by means of a multivariate nonlinear predictor~\cite{kantz1997nonlinear} we are able to achieve a very high degree of accuracy in forecasting future user geographic locations in terms of longitude and latitude. We compare it with traditional linear prediction techniques (such as ARMA~\cite{Cha04})  and we show that these are not able to capture the dynamics of individuals in the geographic space.
\item We discuss how the concept of mutual information can be used to quantify the correlation between two mobility traces and we demonstrate that it is possible to exploit movement data of friends and acquaintances, when such information is available.%, in order to improve movement prediction. These social ties are measured using different indicators: presence in the address book of a user, colocation and number of phone calls.
\item Finally, we study how the correlation measured through mutual information of mobility traces of two individuals, can be used to improve human prediction movement dramatically, also discussing the correlation between human mobility and social ties.
\end{itemize}

The key findings of our analysis are the following: 1) mobility correlation
%and
%the presence of social ties
can be used to improve movement forecasting by
exploiting mobility data of friends; 2) correlated movement is linked to the
existence of physical or distant social interactions and vice versa.

We evaluate these findings on two datasets. The first dataset, which was provided for the Nokia Mobility Data Challenge (NMDC), contains information
related to 39 users~\cite{NokiaMDC}, including the following: GPS traces,
telephone numbers, call and SMS history, Bluetooth and WLAN history. We use the
information of 25 of them, since the dataset does not include phone numbers for
14 of them; therefore, it is not possible to detect if and when phone calls
occur between them. We use GPS traces to analyse the movement of the users.

The second dataset we analyse is Cabspotting~\cite{cabspotting_dataset}, containing mobility traces of
about 500 taxis driving around San Francisco for 30 days. We restrict our
analysis to the 178 taxis with mobility traces longer than 25000 GPS readings. For this dataset we have no
information about relations between taxi drivers (such as friendship connections or co-affiliation).

The paper is organised as follows. In Section~\ref{sec:prediction} we firstly introduce multivariate nonlinear time series prediction techniques and their application to our datasets. Then, in Section~\ref{sec:mutualinformation} we discuss how mutual information can be used to measure the correlation between the movement of two users. Section~\ref{sec:results} focusses on the analysis of the performance improvement that is possible to obtain by considering the traces of highly correlated users as inputs of the predictors. In Section~\ref{sec:discussion}, we discuss our findings also outlining some future directions. Section~\ref{sec:conclusions} concludes the paper, summarising our contributions.

\section{Multivariate Nonlinear Time Series Prediction}
\label{sec:prediction}

\begin{figure}[h]
\center
\includegraphics[width=\columnwidth,clip,trim=2mm 5mm 5mm 2mm]{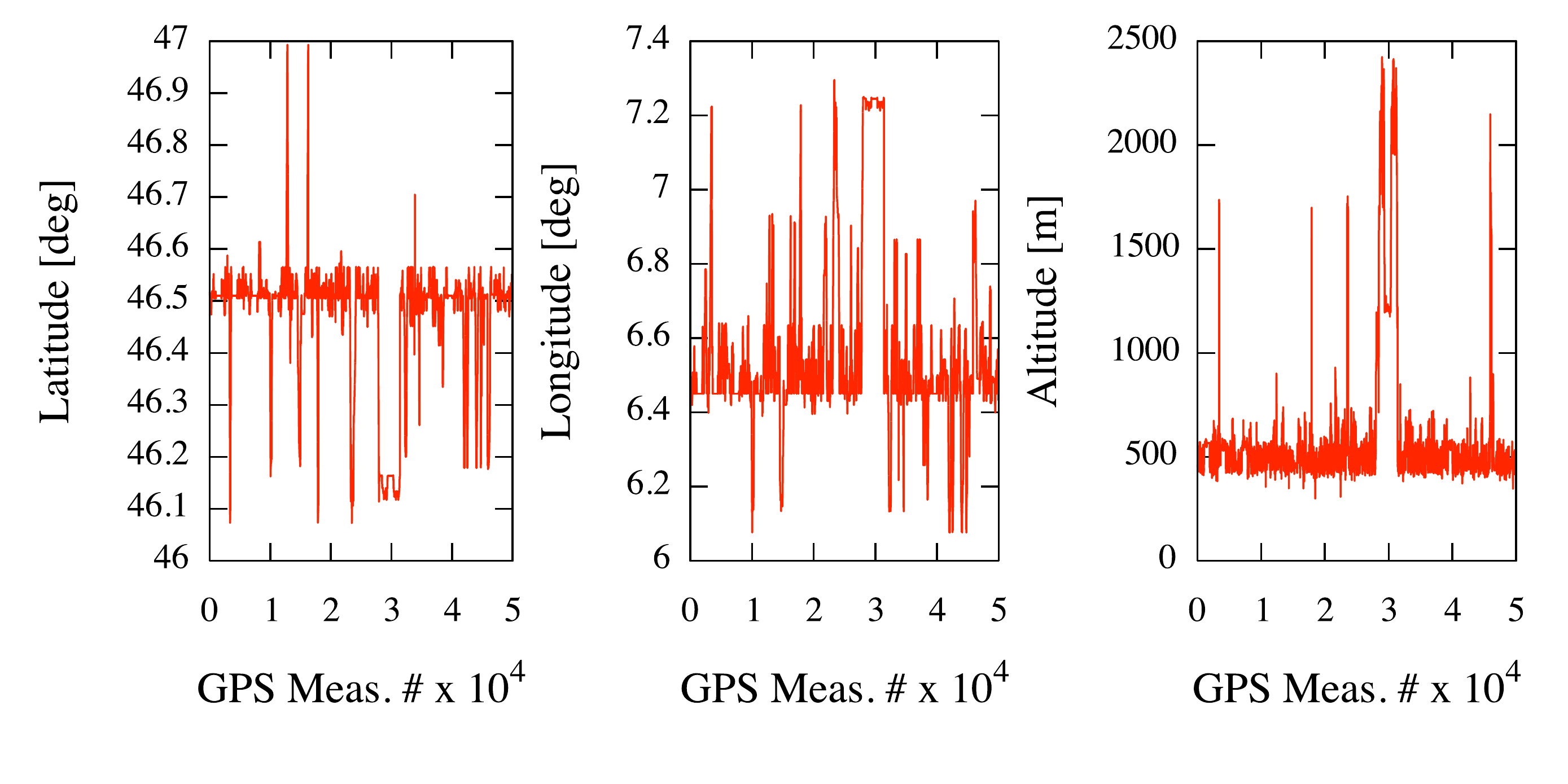}
\caption{Time series from the Nokia Mobility Data Challenge dataset, corresponding to the movements of user 179. No periodic behaviour is apparent in the movement traces of the user.}
\label{fig:time-series-179}
\end{figure}

\begin{figure}[h]
\center
\includegraphics[width=\columnwidth,clip,trim=2mm 5mm 5mm 2mm]{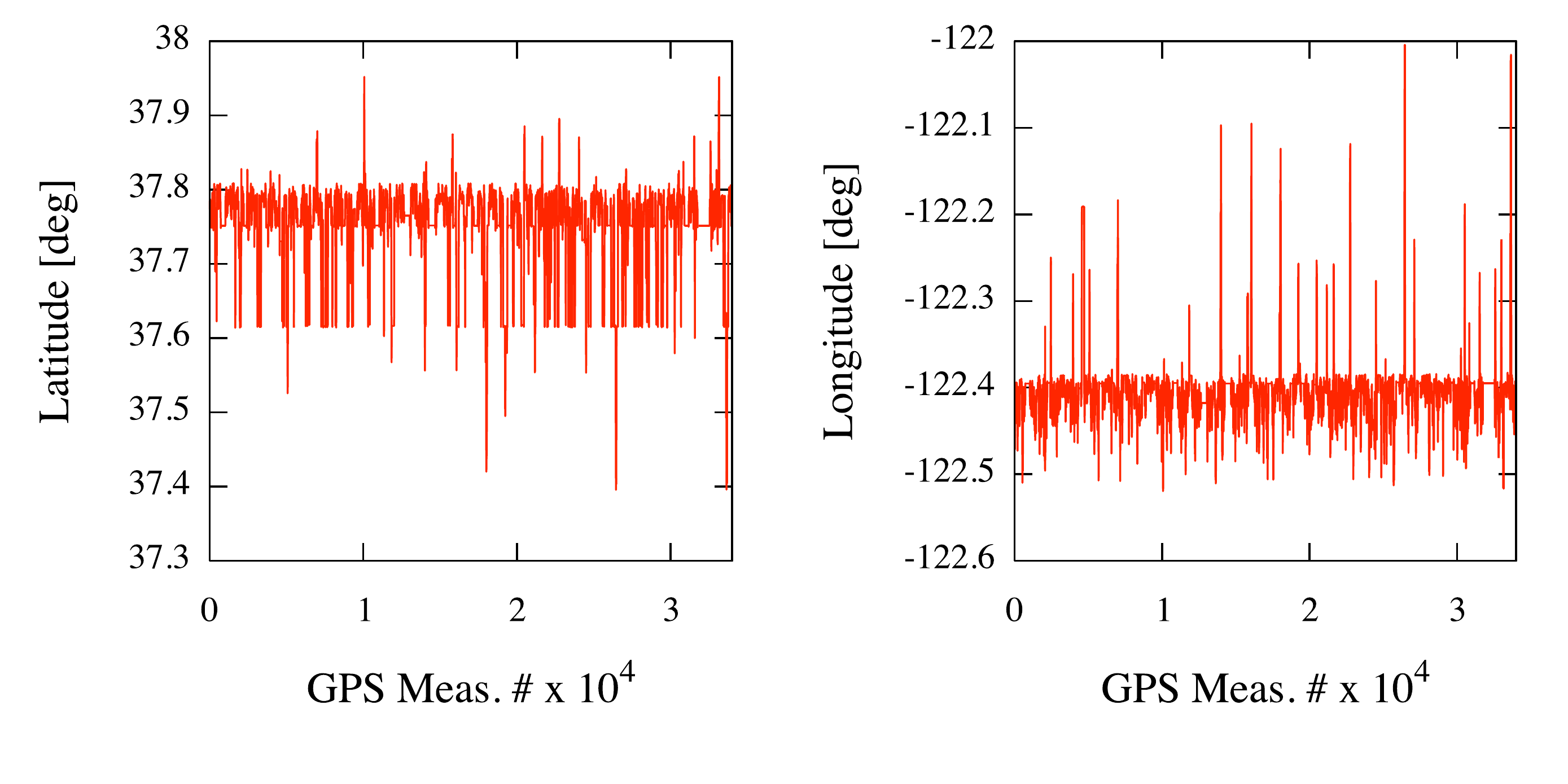}
\caption{Time series from the Cabspotting dataset, corresponding to the movements of taxi \textit{abgibo}.}
\label{fig:time-series-abgibo}
\end{figure}

We now present how we apply nonlinear time series prediction methods to the problem of forecasting the future GPS coordinates of the users, given the past movement history as an input. We will then extend this model by considering also the movement of other users (in particular friends, in the case of the NMDC dataset) as input of the nonlinear predictor. 

\subsection{Overview}
We model the position of a user on the Earth with a time-varying four-dimensional state vector $\mathbf{s}_{n}$ with the following dimensions: hour of the day $h_{n}$, latitude $\phi_{n}$, longitude $\lambda_{n}$ and altitude\footnotemark\footnotetext{The corresponding time series is available only in the NMDC dataset: in the case of Cabspotting data we use a time-varying three-dimensional state vector.} $\xi_{n}$. The prediction of the future states of vector $\mathbf{s}_{n}$ can be performed using different approaches~\cite{kantz1997nonlinear}. We choose the method based on the reconstruction of the phase space of $\mathbf{s}_{n}$ by means of the delay embedding theorem, since this is considered the best state-of-the-art solution to this problem. While the scalar sequence of coordinates may appear completely non deterministic, it is possible to uncover the characteristics
of its dynamic evolution by analysing sub-sequences of the time series itself. In order to investigate the structure of the original system, the time series values must be transformed in a sequence of vectors with a technique called delay embedding.
For a univariate time series measurement $x_{n}$ of a $d-$dimensional dynamical system, the Takens' embedding theorem \cite{takens1981detecting} allows to reconstruct a $m-$dimensional space ($m\geq 2d+1$) with the same dynamical characteristics of the original phase space. The key idea is to build a delay vector $\mathbf{x}_{n}$ by using delayed measurement defined as follows:
\begin{equation}
\mathbf{x}_{n}\equiv(x_{n-(m-1)\tau},x_{n-(m-2)\tau},...,x_{n-\tau},x_{n}),
\end{equation}
where $\tau$ is a time delay. Hence, the reconstruction depends on the two parameters $m$ and $\tau$, which have to be estimated. This technique can be extended to the case of the embedding of a multivariate time series\footnotemark\footnotetext{We refer to \cite{vlachos2009state} (and references therein) for an overview of practical applications of multivariate embedding.} \cite{cao1998dynamics}. 

\begin{figure}[t]
\center
\includegraphics[width=\columnwidth,clip,trim=2mm 5mm 5mm 2mm]{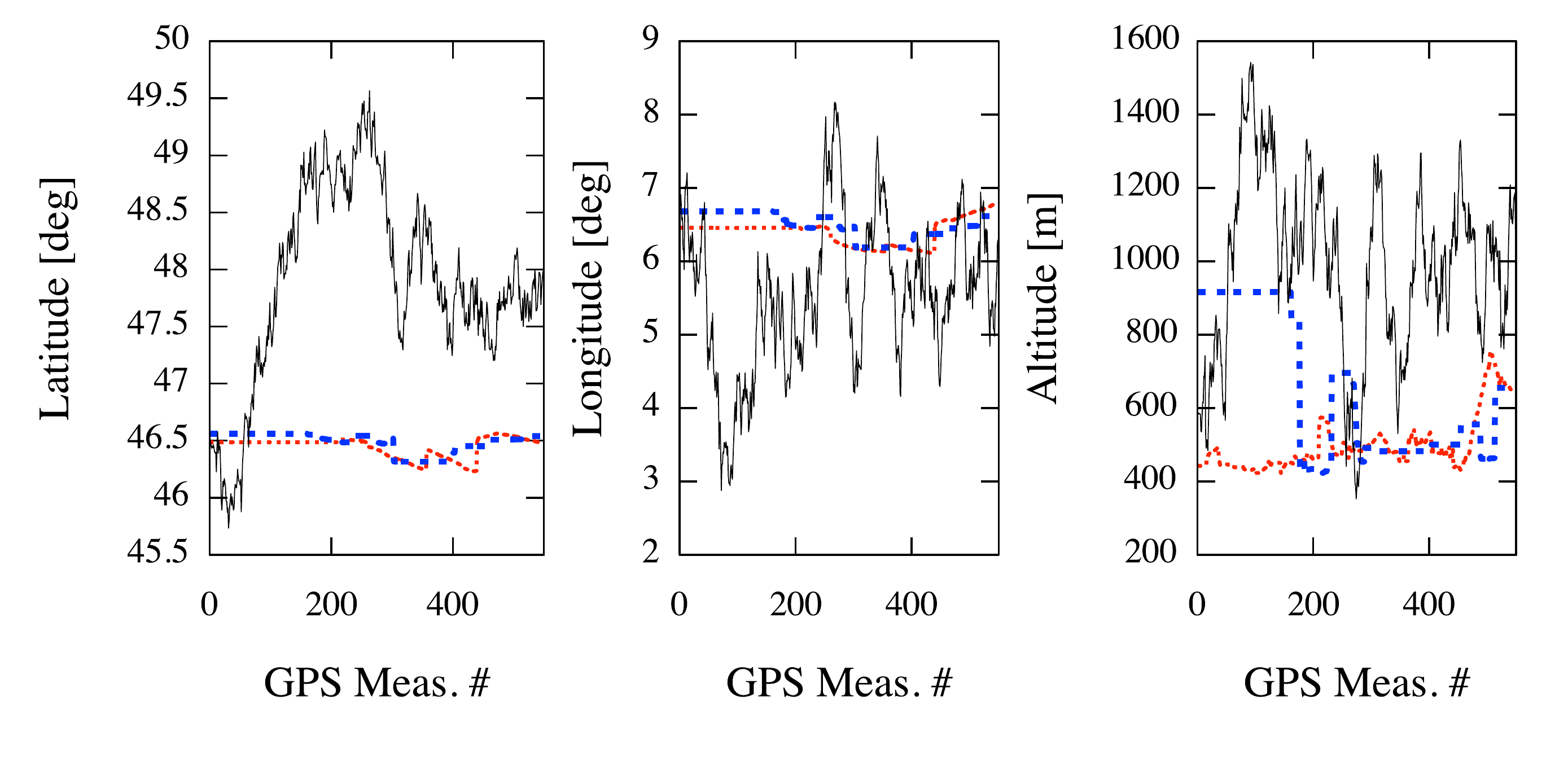}
\caption{Multivariate nonlinear prediction of user 179 mobility in the geographic space: the first 600 predictions, corresponding to about 60 hours, are shown. The dotted line represents the true movements, the solid line indicates the prediction within an ARMA model, while the dashed line indicates the data obtained by means of a multivariate nonlinear predictor.}
\label{fig:pred-179}
\end{figure}

Under the hypotheses of Takens' theorem, i.e., non-noisy time
series of infinite length, the underlying dynamics can be fully reconstructed
by using only univariate measurements of the dynamical system of interest.
Unfortunately, real-world measurement are noisy and with finite length: hence,
the phase space reconstruction is more precise if multivariate measurements of
the dynamical system under investigation are performed.

\fussy Let us indicate with $N$ the number of measurements corresponding to an $M-$dimensional time series $\mathbf{y}_{1}$, $\mathbf{y}_{2}$, ..., $\mathbf{y}_{N}$, with $\mathbf{y}_{i}\equiv(y_{1,i},y_{2,i},...,y_{M,i})$ and $i=1,2,...,N$. The resulting delay vector is
\begin{eqnarray}
\mathbf{v}_{n}&\equiv&(y_{1,n-(m_{1}-1)\tau_{1}},y_{1,n-(m_{1}-2)\tau_{1}},...,y_{1,n},\nonumber\\
&&y_{2,n-(m_{2}-1)\tau_{2}},y_{2,n-(m_{2}-2)\tau_{2}},...,y_{2,n},\nonumber\\
&&...\nonumber\\
&&y_{M,n-(m_{M}-1)\tau_{M}},y_{1,n-(m_{M}-2)\tau_{M}},...,y_{M,n}),
\end{eqnarray}
where $m_{j}$ and $\tau_{j}$, $j=1,2,...,M$ are respectively the embedding and time delays corresponding to each component of the multivariate time series.

Intuitively, this method searches the past history to find and extract sequences of values that are
very similar to the recent history. Assuming a certain degree of determinism in the system, the assumption is that, given a certain state (in our case geographic coordinates), there is a strong probability that this will be followed by the same next state.

\subsection{Evaluation}

\subsubsection{Linearity Analysis}

The complexity of the time series taken into consideration in our study is apparent by observing the two representative examples shown in Fig.\,\ref{fig:time-series-179} and Fig.\,\ref{fig:time-series-abgibo}. The figures show thousands of time-ordered GPS measurements corresponding to the position on the Earth of user 179 (NMDC dataset) and taxi \textit{abgibo} (Cabspotting dataset), respectively.

We firstly apply linear prediction models to these time series. The time series appear rather noisy with alternating spikes, nearly flat values, corresponding to stationary points, and fluctuation around an average value. We try to model such movements in the space with a simple multivariate AR$(p)$~+~noise process. 

As for the order $p$ of the multivariate autoregressive model that best approximates the original time series, we choose the one that minimises an information criterion, according to Akaike \cite{akaike1974new} and Schwarz \cite{schwarz1978estimating}. We find that $p=24$ provides the best approximation. Hence, we use such a model to perform a multivariate linear forecasting of 1000 GPS measurements for user 179 (NMDC dataset). We validate the model by comparing the latest 1000 real GPS measurements against the forecasted ones\footnotemark\footnotetext{The latest 1000 real GPS measurements have not been included in the procedure adopted to estimate the best order $p$.}. The results are shown in Fig.\,\ref{fig:pred-179}, where the real movements are indicated with dots and the forecasting with the linear model is indicated by the solid line. It is evident that the forecasting is not in agreement with observations. In fact, the prediction error on the position (latitude and longitude) is of the order of $3^{\circ}$, whereas the error on the altitude is generally larger than 600 m.

However, although the time series are not regularly sampled, we find that they show some features typical of deterministic dynamics contaminated by noise. In fact, preliminary inspection of phase space reconstruction by means of Takens' embedding theorem shows an underlying structure, typical of deterministic dynamical systems. This aspect will be addressed more quantitatively in the remainder of the paper.

\subsubsection{Estimation of the Embedding Dimension and Time Delay}
\begin{figure}[tbh]
\center
\includegraphics[width=\columnwidth,clip,trim=2mm 5mm 5mm 2mm]{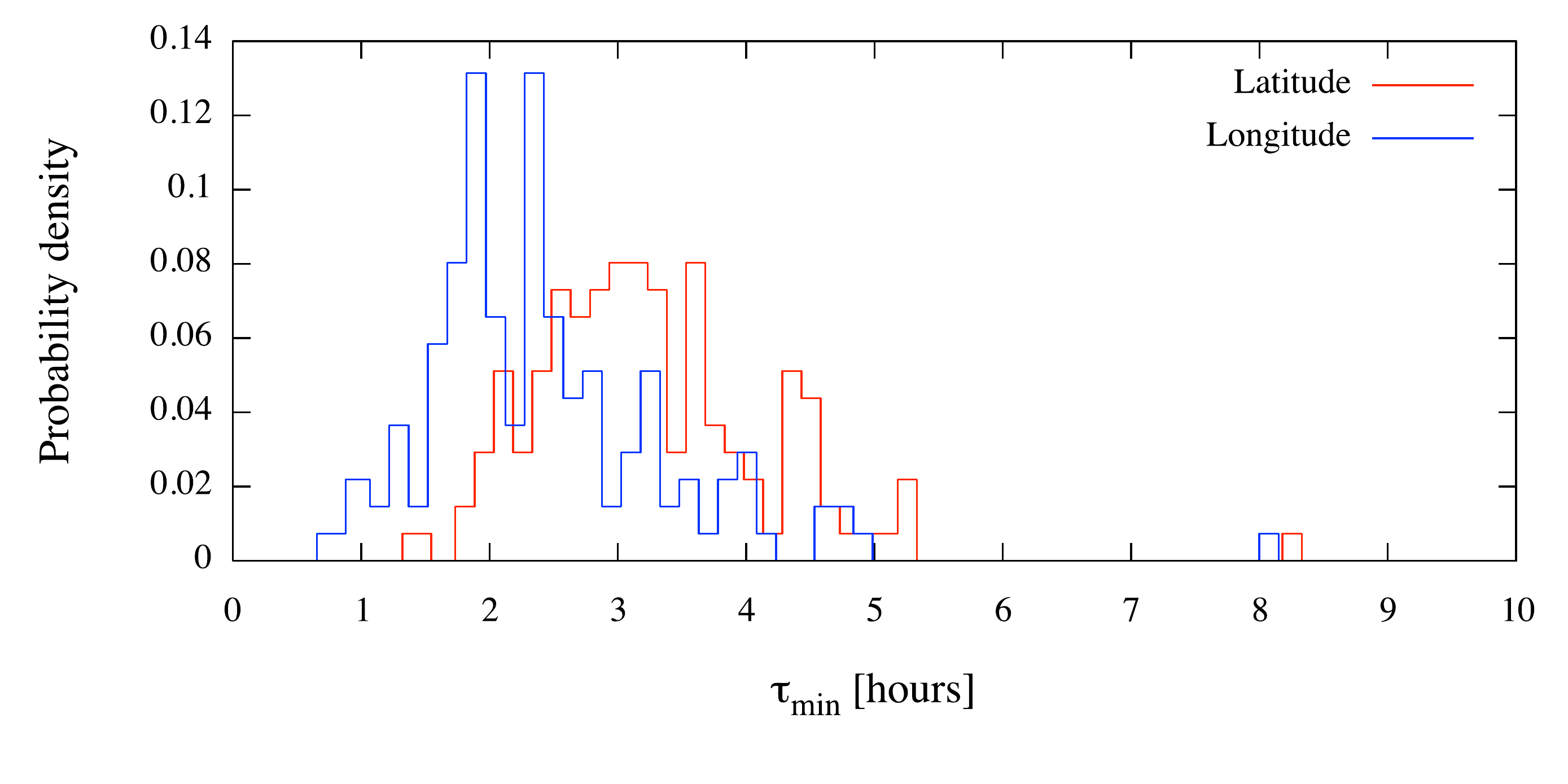}
\caption{\emph{Cabspotting dataset.} Distribution of the time delay $\tau_{min}$ corresponding to the first local minimum of the average mutual information obtained from the time series of latitude and longitude. The average value of $\tau_{min}$ is approximately 3 hours.}
\label{fig:amin-cab}
\end{figure}

Although several methods have been proposed to estimate the values of embedding and time delay, in our analysis we consider the same time delay $\tau$ for all the series. In fact, for a given user, we have found that the time delay $\tau_{min}$ corresponding to the first local minimum of the average mutual information \cite{fraser1986independent}, generally adopted to estimate $\tau$ in the univariate case, is of the same order of magnitude for any component. As a representative example, in Fig.\,\ref{fig:amin-cab} we show the distribution of $\tau_{min}$ obtained from the time series of latitude and longitude of taxis in the Cabspotting dataset.

\begin{figure}[tbh]
\center
\includegraphics[width=\columnwidth,clip,trim=2mm 3mm 3mm 2mm]{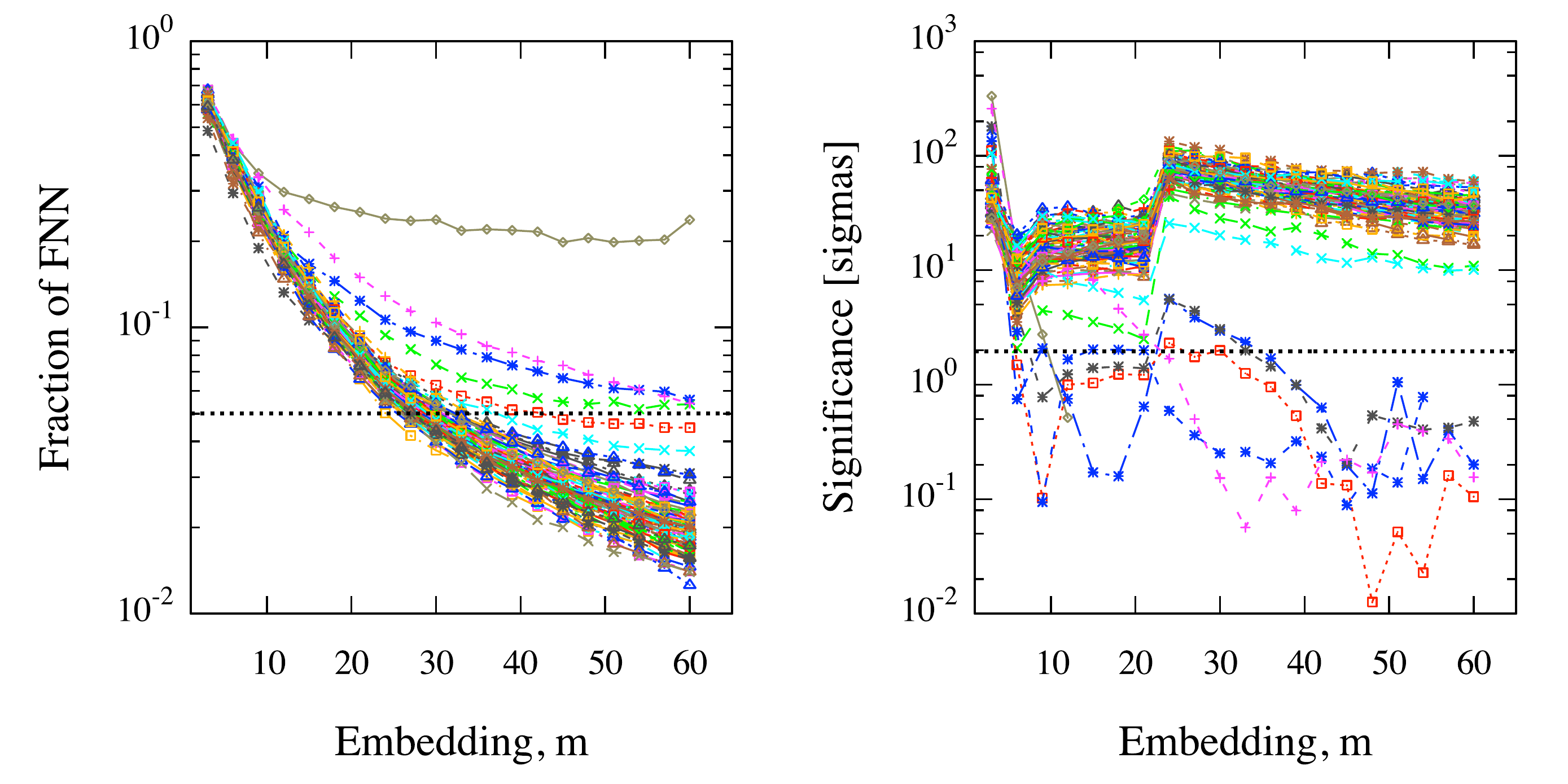}
\caption{\emph{Cabspotting dataset.} Left panel: fraction of false nearest neighbours as a function of embedding dimension $m$; each curve corresponds to a different mobility trace and the threshold chosen to estimate the optimal embedding is indicated by the dotted horizontal line. Right panel: significance in rejecting the null hypothesis by using the fraction of false nearest neighbours as discriminator in the surrogate test (see the text for further details); each curve corresponds to a different mobility trace and the size of the statistical test is indicated by the dotted horizontal line.}
\label{fig:fnn-cab}
\end{figure}

This fact has also practical implications, since it simplifies the application of this methodology for the analysis of our data. The optimal embedding dimension is estimated by exploiting the method of false nearest neighbours \cite{kennel1992determining,kantz1997nonlinear,hegger1999improved} in the case of multivariate embedding \cite{boccaletti2002reconstructing}. For any point in the data, an $m^{\star}$-dimensional phase space is considered and the number of false nearest neighbours, i.e., points which are neighbours in the $m^{\star}-$dimensional space but not in the $(m^{\star}+1)-$dimensional one, is estimated. The desirable embedding dimension $m$ is such that the percentage of false nearest neighbours is small, e.g., below 5\%. Any efficient algorithm for counting nearest neighbours is allowed: in particular, we adopt the method implemented in the TISEAN software \cite{hegger1999practical}. In the left panel of Fig.\,\ref{fig:fnn-cab} we show the fraction of false nearest neighbours as a function of $m$, obtained from mobility traces in the Cabspotting dataset. For any trace, the optimal embedding dimension is close to 30, confirming that the underlying dynamics is very similar\footnotemark\footnotetext{We find only a few exceptions whose number represents less than 5\% of the mobility traces in the whole dataset.}. The false nearest neighbour method alone is not able to distinguish between deterministic and stochastic processes on an absolute level \cite{hegger1999improved}: however, it is among the state-of-the-art solutions \,\cite{Farmer87,Sugihara90,Barahona96,Schmitz97,dedomenico2010fast} that can be reliably used to asses the nonlinearity of time series by means of a statistical test with surrogate data.

\subsubsection{Analysis of Multivariate Surrogates}
Given a multivariate time series $\{\mathbf{s}_{n}\}$, we produce a set $\{\hat{\mathbf{s}}^{(i)}_{n}\}$, $i=1,2,...,N$, of
$N$ multivariate surrogates of $\{\mathbf{s}_{n}\}$. The surrogates are synthetic time series, built from $\{\mathbf{s}_{n}\}$, preserving both statistical and linear features of the original time series, as probability distribution and autocorrelation, while removing the effects of nonlinearities and nonstationarities, if any. In particular, we adopt the iterative amplitude-adjusting Fourier transform (IAAFT) scheme\,\cite{Schr-Schm96,Schreiber98} to build surrogates. Hence, we choose the fraction of false nearest neighbours as discriminator to test the null hypothesis that the mobility traces can be described by a linear stochastic model. Let us indicate with $q(m)$ the value of the discriminator obtained for an embedding dimension $m$ from the observed multivariate time series, and with $\hat{q}^{(i)}(m)$ the values of the discriminator obtained from surrogates. Our numerical experiments indicate that the distribution of $\hat{q}^{(i)}(m)$ is described with a reasonable approximation by a Gaussian function with average $\mu_{\hat{q}}(m)$ and variance $\sigma_{\hat{q}}^{2}(m)$. This fact allows us to define the quantity 
\begin{eqnarray}
\Sigma(m)\equiv\frac{|q(m)-\mu_{\hat{q}}(m)|}{\sigma_{\hat{q}}(m)}
\end{eqnarray}
as a measure of \textit{significance}. In this case, if the null hypothesis is true then the $p-$value of observing a significance equal or larger than $\Sigma(m)$ is given by $p(m)=\textrm{erfc}[\Sigma(m)/\sqrt{2})]$. We fix \emph{a priori} the size $\alpha=0.05$ of our hypothesis testing: if $p(m)<\alpha$ (or, equivalently, if $\Sigma(m)>1.96$) the null hypothesis that mobility can be described by a linear stochastic model is rejected with 95\% confidence level (CL). In the right panel of Fig.\,\ref{fig:fnn-cab} we show the significance as a function of $m$ for mobility traces in the Cabspotting dataset. Remarkably, the significance is much larger than 1.96 for all traces despite a few exceptions, independently from the embedding dimension chosen for the reconstruction. Hence, we can conclude that human mobility exhibits a strong nonlinear dynamics. Moreover, the existence of short-term correlations, as indicated by the average mutual information analysis and of decreasing fraction of false nearest neighbours for increasing embedding dimension suggests that such a dynamics should have a deterministic component potentially contaminated by a stochastic dynamics.

\subsubsection{Analysis of Prediction Errors}
Dealing with nonlinear dynamical systems with a potential deterministic component, we adopt a method which exploits such features to predict the future movements of users in the NMDC dataset and of taxis in the Cabspotting dataset. The multivariate nonlinear prediction (MNP) is performed by approximating the dynamics locally in the phase space by a constant (see~\cite{casdagli1989nonlinear} for further information). In the delay embedding space, all the points in the neighbourhood $\mathcal{U}_{n}$ of the state $\mathbf{v}_{n}$ are taken into account in order to predict the coordinates at time $n+k$. Hence, the forecast $\hat{\mathbf{v}}_{n+k}$ for $\mathbf{v}_{n+k}$ is given by  
\begin{eqnarray}
\hat{\mathbf{v}}_{n+k}=\frac{1}{|\mathcal{U}_{n}|}\sum_{\mathbf{v}_{j}\in\mathcal{U}_{n}}\mathbf{v}_{j+k},
\end{eqnarray}
i.e., the average over the states which correspond to measurements $k$ steps ahead of the neighbours $\mathbf{v}_{j}$.

Hence, we use MNP to forecast the same 1000 GPS measurements previously discussed in the case of NMDC dataset. Again, we validate the model by comparing the latest 1000 real GPS measurements against the forecasted ones. The results for user 179 are shown in Fig.\,\ref{fig:pred-179}, where the real movements are indicated with triangles and the forecasting with the nonlinear method is indicated by the dashed line. The number of nearest neighbours used to build the neighbourhood $\mathcal{U}_{n}$ has been kept fixed to 10. Intriguingly, the nonlinear forecasting is in excellent agreement with observations of latitude and longitude, with a global position prediction error equal to $0.19^{\circ}$, and in good agreement with the altitude coordinate, with a global altitude forecasting error equal to 219.43 m.

The global error on the time series prediction is estimated separately for each component using the following formula:
\begin{eqnarray}
\label{def:rmse}
e_{j}=\sqrt{\frac{1}{N}\sum_{n=1}^{N}(\hat{s}_{j,n}-s_{j,n})^{2}},
\end{eqnarray}
with $j=1,2,...,M$ with $M=4$, $N=1000$. The overall error between the predicted position and the real one is given by the geodesic distance.

% Nel caso specifico di lat e long ($M=2$ e $M=3$, rispettivamente), l'errore totale è $e=\sqrt{e_{2}^{2}+e_{3}^{2}}$. Va trovato un modo per introdurre questo , se vogliamo e abbiamo spazio, magari anche come footnotemark}@

%Fig.\,\ref{fig:time-series-179} 

%Fig.\,\ref{fig:pred-179}

\section{Mutual Information and Movement Correlation}
\label{sec:mutualinformation}
In this section, we will briefly introduce the concept of mutual information
and we will show how this quantity can be exploited in our analysis to measure
the correlation between the movement of different individuals. In the following
section, we will then discuss how mutual information can be used to select
mobility data of other users that can be exploited as inputs of the nonlinear
predictors in order to improve the prediction accuracy.

\subsection{Overview}

Let us assume that $\mathbf{X}$ and $\mathbf{Y}$ are two multivariate
stochastic variables, and let us indicate with $P_{\mathbf{X}}(\mathbf{x})$ and
$P_{\mathbf{Y}}(\mathbf{y})$, respectively, the corresponding Probability
Density Functions (PDF). The joint probability is indicated by
$P_{\mathbf{X}\mathbf{Y}}(\mathbf{x},\mathbf{y})$. The mutual information
$\mathcal{I}(\mathbf{X},\mathbf{Y})$ between such two variables is defined as
follows:
\begin{eqnarray}
\label{def:mi}
\mathcal{I}(\mathbf{X},\mathbf{Y})=\sum_{\mathbf{x}\in\mathbf{X}}\sum_{\mathbf{y}\in\mathbf{Y}} P_{\mathbf{X}\mathbf{Y}}(\mathbf{x},\mathbf{y})\log \frac{P_{\mathbf{X}\mathbf{Y}}(\mathbf{x},\mathbf{y})}{P_{\mathbf{X}}(\mathbf{x})P_{\mathbf{Y}}(\mathbf{y})}.
\end{eqnarray}
The mutual information\footnotemark\footnotetext{The units of mutual information are nats when the natural logarithm is used.} quantifies how much information the variable $\mathbf{Y}$ provides about the variable $\mathbf{X}$. For this reason, it can be used as an estimator of the amount of correlation between $\mathbf{X}$ and $\mathbf{Y}$. In fact, if the two variables are totally uncorrelated then $P_{\mathbf{X}\mathbf{Y}}(\mathbf{x},\mathbf{y})=P_{\mathbf{X}}(\mathbf{x})P_{\mathbf{Y}}(\mathbf{y})$ and $\mathcal{I}(\mathbf{X},\mathbf{Y})=0$.

%\footnote{It is interesting to note that mutual information is strictly related to the entropy of the two series. In fact we can write the expression of mutual information as follows:
%\begin{eqnarray}
%\mathcal{I}(\mathbf{X},\mathbf{Y})&=&H(\mathbf{X})-H(\mathbf{X}|\mathbf{Y})\nonumber\\
%&=&H(\mathbf{Y})-H(\mathbf{Y}|\mathbf{X})\nonumber\\
%&=&H(\mathbf{X},\mathbf{Y})-H(\mathbf{X}|\mathbf{Y})-H(\mathbf{Y}|\mathbf{X}),\nonumber
%\end{eqnarray}
%being $H(\mathbf{X})$ the information entropy of $\mathbf{X}$, $H(\mathbf{X}|\mathbf{Y})$ the conditional information entropy of $\mathbf{X}$ with respect to $\mathbf{Y}$ and the joint information entropy $H(\mathbf{X},\mathbf{Y})$.}.

%\mm{footnote about mutual information and entropy can be cut if necessary}

In our analysis $\mathbf{X}$ represents the motion of a user on the Earth, the
random samples $\mathbf{x}$ drawn from $\mathbf{X}$ correspond to geographic
coordinates, whereas the PDF of $\mathbf{x}$ quantifies the fraction of time
spent by the user in a particular position.

We use the mutual information to quantify the amount of correlation between the
motion of different users, or, equivalently, how much information the motion
$\mathbf{Y}$ provides about the motion $\mathbf{X}$. 

\subsection{Evaluation}

\begin{figure}[tbh]
\center
\includegraphics[width=\columnwidth,clip,trim=2mm 5mm 5mm 2mm]{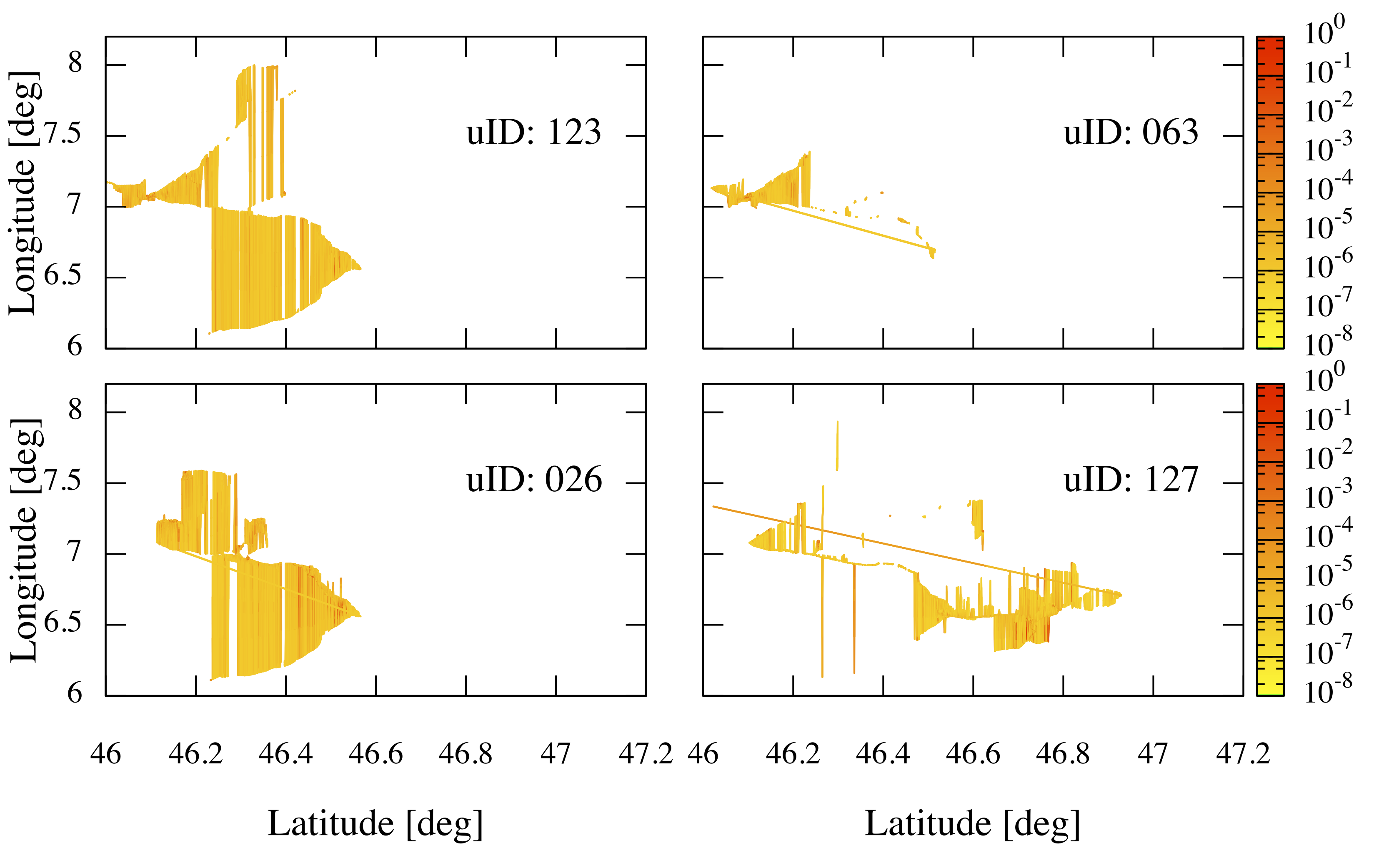}

\caption{\emph{NMDC dataset.} PDF of locations occupied by four different users. Top: users are
friends or acquaintances. We say that two individuals are friends or
acquaintances if one of them is in the other's address book. Bottom:
users are not friends or acquaintances. The colour indicates the frequency of
occupation.}
\label{fig:gps-density}
\end{figure}

In the NMDC dataset, we say that two individuals are friends or acquaintances if one of them is in the other's address book.
In Fig.\,\ref{fig:gps-density} the
two-dimensional PDF of positions occupied by four different users is shown.
Users 063 and 123 are friends or acquaintances, while users 026 and 127 are
not.  

\section{Exploiting Movement Correlation and Social Ties to Improve Prediction Accuracy}
\label{sec:results}
We now discuss how mobility traces of individuals that have correlated
geographic patterns and social ties can be used to improve the accuracy of
movement forecasting. 

\begin{figure}[tbh]
\center
\includegraphics[width=\columnwidth,clip,trim=2mm 5mm 5mm 2mm]{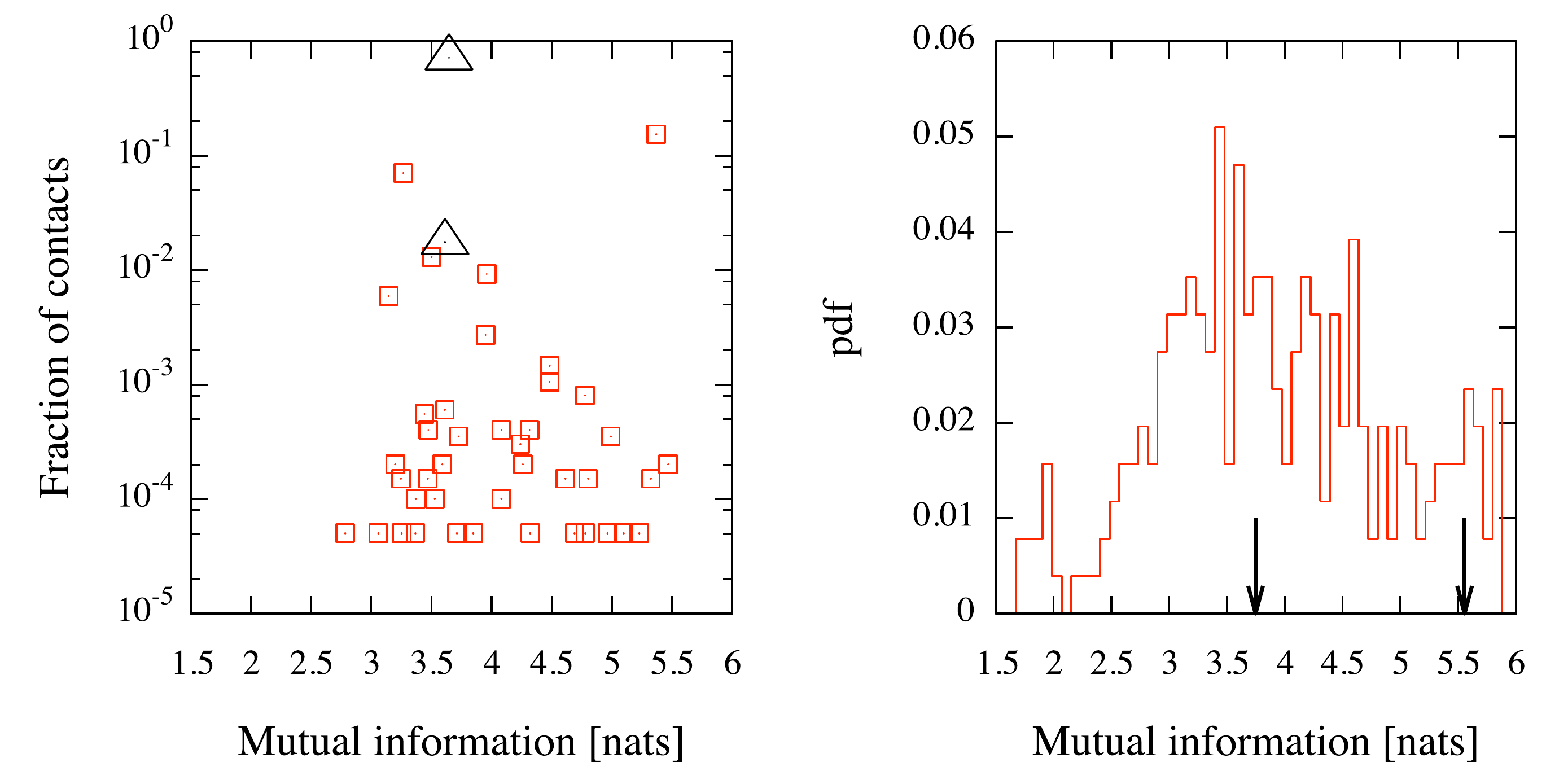}
\caption{\emph{NMDC dataset.} Left panel: scatter plot of the fraction of contacts vs the mutual information estimated for pairs of users with at least one contact, where triangles indicate the two pairs of users connected by social ties in the dataset. Right panel: pdf of mutual information estimated for pairs of users with no contacts at all, where arrows indicate the value of mutual information for the only two pairs of user with social ties, and no contacts, in the dataset.}
\label{fig:aggr-vs-mi}
\end{figure}

\subsection{Our Approach at a Glance}

Our approach can be summarised as follows: assuming that we want to predict the
movement of person/entity $A$, instead of having only the vector describing the location of
$A$ as input, we will also consider the movement history of another person/entity $B$,
characterised by mobility patterns that are strongly correlated to those of the
user we would like to predict. This measure is given by the mutual information
introduced in the previous section. 
%$A$ and $B$ are users or taxis depending on
%the dataset we are considering.

From a mathematical point of view, the idea is to use a 8-dimensional vector
that is given by the juxtaposition of the two time-varying state vectors
representing the states (time-stamped GPS coordinates) of $A$ and 
$B$, which we indicate with 
$\mathbf{s}_{n_A}$ and $\mathbf{s}_{n_B}$, as inputs of the multivariate
nonlinear predictor.

\subsection{Evaluation}
In both datasets we find that by using additional traces of pairs with high
correlation, the accuracy of the prediction improves consistently. In the case of NMDC, 
the improvement is of at least one order of magnitude (and often of two orders of
magnitude) with respect to the prediction based on only single traces.  Moreover, it is
interesting to note that social ties can also be used to select the user for
the additional traces as input. In fact, we find that if we select mobility
patterns of individuals that are in the address book of the user, the
performance of the predictor improves dramatically.  At the same time, we would
like to stress the fact that the NMDC dataset contains a small number of users, therefore
it is difficult to make claims about the general validity of these findings.
However, we find the same results for much larger Cabspotting dataset. In this dataset, it is not possible to use social ties\footnote{In theory, it might be interesting to investigate the influence of the social ties between taxi drivers, but this information unfortunately is not available in the dataset.}, but we find that 
if we select mobility patterns of taxis whose mutual information is high, the
performance of the predictor improves drastically.

Hence, we perform the same analysis described in Section~\ref{sec:prediction},
but including the time series of
movements corresponding to other users in the multivariate nonlinear prediction.
%In particular, we use three pairs of users with: i) no social ties; ii)
%unidirectional ties and iii) bidirectional ties.
The global prediction error, defined by Eq.\,(\ref{def:rmse}), of position and
altitude is reported in Tab.\,\ref{tab:err} for three pairs in the NMDC dataset. As shown in
this table, we observe that the additional information provided by the movement
of a user socially linked to that taken into consideration improves the
prediction by more than one order of magnitude with respect to the case of
users who are not socially linked to each other. 
%However,  we also note that we do not claim that this is a general result,
%given the limited size of the dataset taken into consideration.

\begin{table}[tbh]
\vspace{5pt}
\begin{center}
\begin{tabular}{| c | c | c | c |}
\hline
Nodes & Social link & Position Error & Altitude Error\\
\hline \hline
026 127 & None & $0.167^{\circ}$ & 66.33 m\\
\hline
063 123 & Present & $0.011^{\circ}$ & 20.95 m\\
\hline
094 009 & Present & $0.003^{\circ}$ & 5.57 m\\
\hline
\end{tabular}
\end{center}
\caption{\emph{NMDC dataset.} Global error, defined by Eq.\,(\ref{def:rmse}), on the prediction of
position and altitude for pairs of users connected through social links
(defined as presence in the address book of the user).} \label{tab:err}
\end{table}

For each pair of users in the NMDC dataset, we count the total number of
Bluetooth contacts and calls, then we estimate their mutual information defined
by Eq.\,(\ref{def:mi}). In order to quantify the amount of correlation between
the fraction of contacts and the mutual information, we build a scatter plot
between these two observables. The result is shown in the left panel of
Fig.\,\ref{fig:aggr-vs-mi}, by considering only pair of users with at least one
contact. The points corresponding to pairs of users with social ties are also
shown (triangles).  In the right panel of Fig.\,\ref{fig:aggr-vs-mi}, we show
the PDF of mutual information obtained by considering only pairs of users with
no contacts at all. The mutual information corresponding to pairs of users with
social ties is shown (arrows). Even if these plots show interesting
correlations for this specific dataset, we believe no generalisations can be
drawn from them, because of the lack of sufficient statistics.

\begin{figure}[tbh]
\center
\includegraphics[width=\columnwidth,clip,trim=2mm 5mm 5mm 2mm]{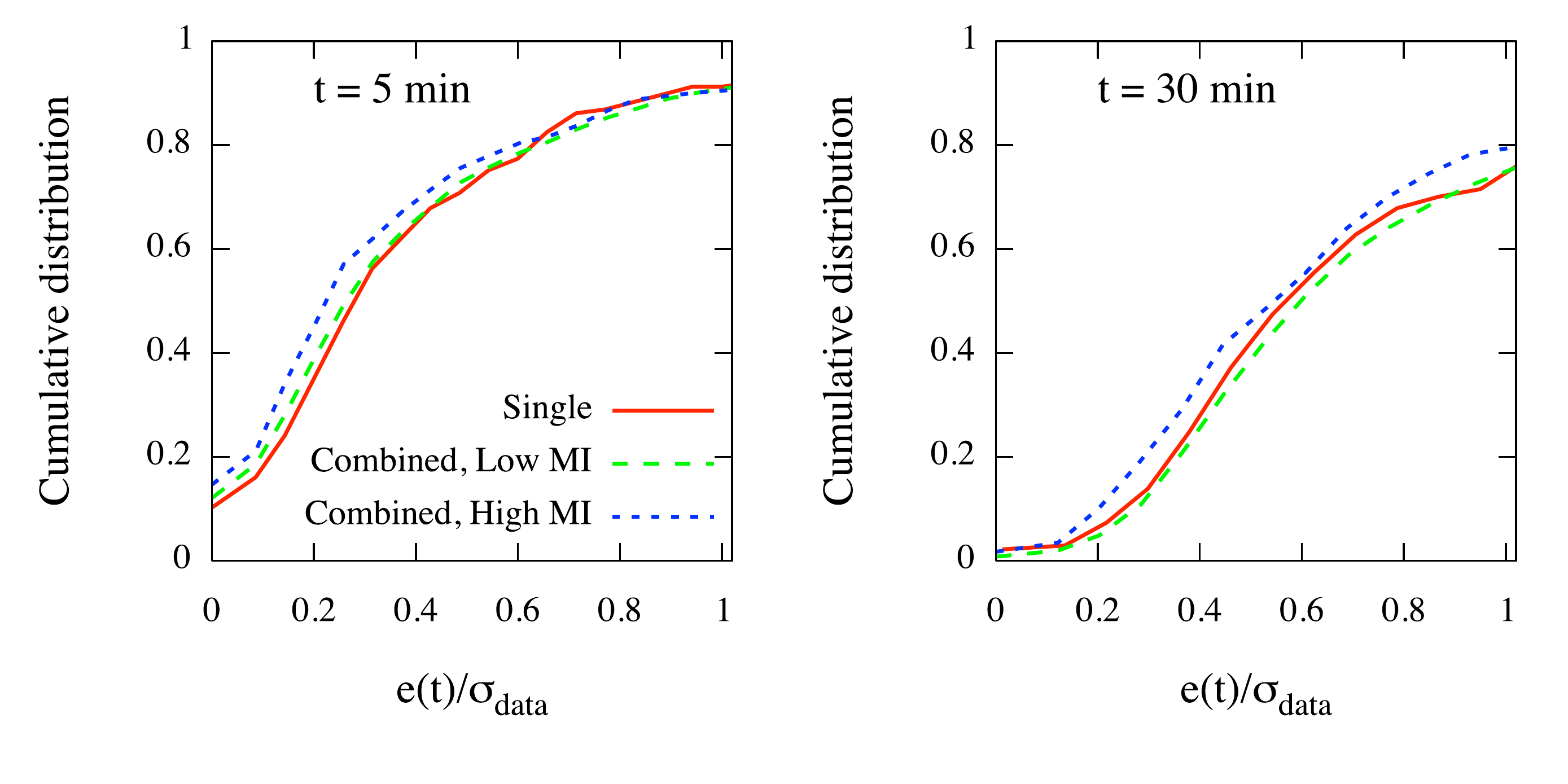}
\caption{\emph{Cabspotting dataset.} Cumulative distribution of the values of the ratio between the global prediction error $e(t)$ and the statistical uncertainty of the mobility traces $\sigma_{data}$. We show the distributions corresponding to the predicted positions after 5 minutes (left panel) and 30 minutes (right panel). The three curves correspond to prediction involving: a) only the past history of each single taxi (``Single''), b) the history of any pairs of taxi whose mobility patterns show a low mutual information (``Combined, Low MI'') and c) the history of any pairs of taxi whose mobility patterns show a high mutual information (``Combined, High MI'').}
\label{fig:prederr-cab}
\end{figure}

Hence, we perform the same analysis by exploiting the mobility patterns of taxis in the
Cabspotting dataset, which contains a larger statistical sample.
In this case, the global prediction error, defined by Eq.\,(\ref{def:rmse}), refers only to 
latitude and longitude. Moreover, we investigate the evolution of the global prediction error
by estimating how it changes versus time. More specifically, we define the time-varying global 
prediction error for each component as
\begin{eqnarray}
\label{def:trmse}
e_{j}(t)=\sqrt{\frac{1}{t}\sum_{n=1}^{t}(\hat{s}_{j,n}-s_{j,n})^{2}},
\end{eqnarray}
with $j=1,2$ and $t$ indicating the prediction interval. Hence, the overall error between the predicted position and the real one at time $t$ is given by $e(t)=\sqrt{e_{1}^{2}(t)+e_{2}^{2}(t)}$. In order to investigate the quality of our prediction, we study the ratio of $e(t)$ with respect to the global statistical uncertainty $\sigma_{data}$ on the position of the taxi. In fact, as long as the ratio $e(t)/\sigma_{data}$ is equal or smaller than one, or, equivalently, if $e(t)\leq\sigma_{data}$, the prediction at time $t$ is within the statistical uncertainty and, therefore, the performance of our predictor can be considered satisfactory. In Fig.\,\ref{fig:prederr-cab} we show the cumulative distribution of the values of the ratio obtained from mobility traces in the Cabspotting dataset. In particular, we show the distributions corresponding to the predicted positions after 5 minutes and 30 minutes. The three curves correspond to prediction involving: a) only the past history of each single taxi (``Single''), b) the history of any pairs of taxi whose mobility patterns show a low mutual information (``Combined, Low MI'') and c) the history of any pairs of taxi whose mobility patterns show a mutual information (``Combined, High MI''). It is worth remarking that the mutual information is not an upper-bounded measure of correlation: hence, we define ``High MI'' all pairs of mobility patterns whose mutual information is distributed among the highest 5\% of values, and ``Low MI'' the remaining pairs of mobility patterns. In both panels of Fig.\,\ref{fig:prederr-cab}, we can observe that the prediction improves when combining pairs of correlated mobility patterns. Moreover, it is intriguing that our method is able to predict in the 80\% of cases the movements of taxis for the next 30 minutes, with an error equal or smaller than the statistical uncertainty of their mobility patterns.

%The complementary cumulative distribution of the absolute values of the global prediction error are shown in Fig.\,\ref{fig:rms-cab} for $t=5$\,minutes (left panel) and $t=30$\,minutes (right panel). The figure indicates the probability of predicting at time $t$ the position of taxis with an error equal or larger than $e(t)$. \ma{Here a) b) and c) appears to behave similarly, but they are not as shown in the previous figure. If you think that this figure might lead to confusion, I suggest to remove it.}
%\al{I agree, for me nothing can be said from this figure, the three distributions are overlapping.}

%\begin{figure}[t]
%\center
%\includegraphics[width=\columnwidth,clip,trim=2mm 5mm 5mm 2mm]{figures/ccdf_rms_comparison.pdf}
%\caption{\emph{Cabspotting dataset.} }
%\label{fig:rms-cab}
%\end{figure}

\begin{figure}[tbh]
\center
\includegraphics[width=\columnwidth,clip,trim=2mm 5mm 5mm 2mm]{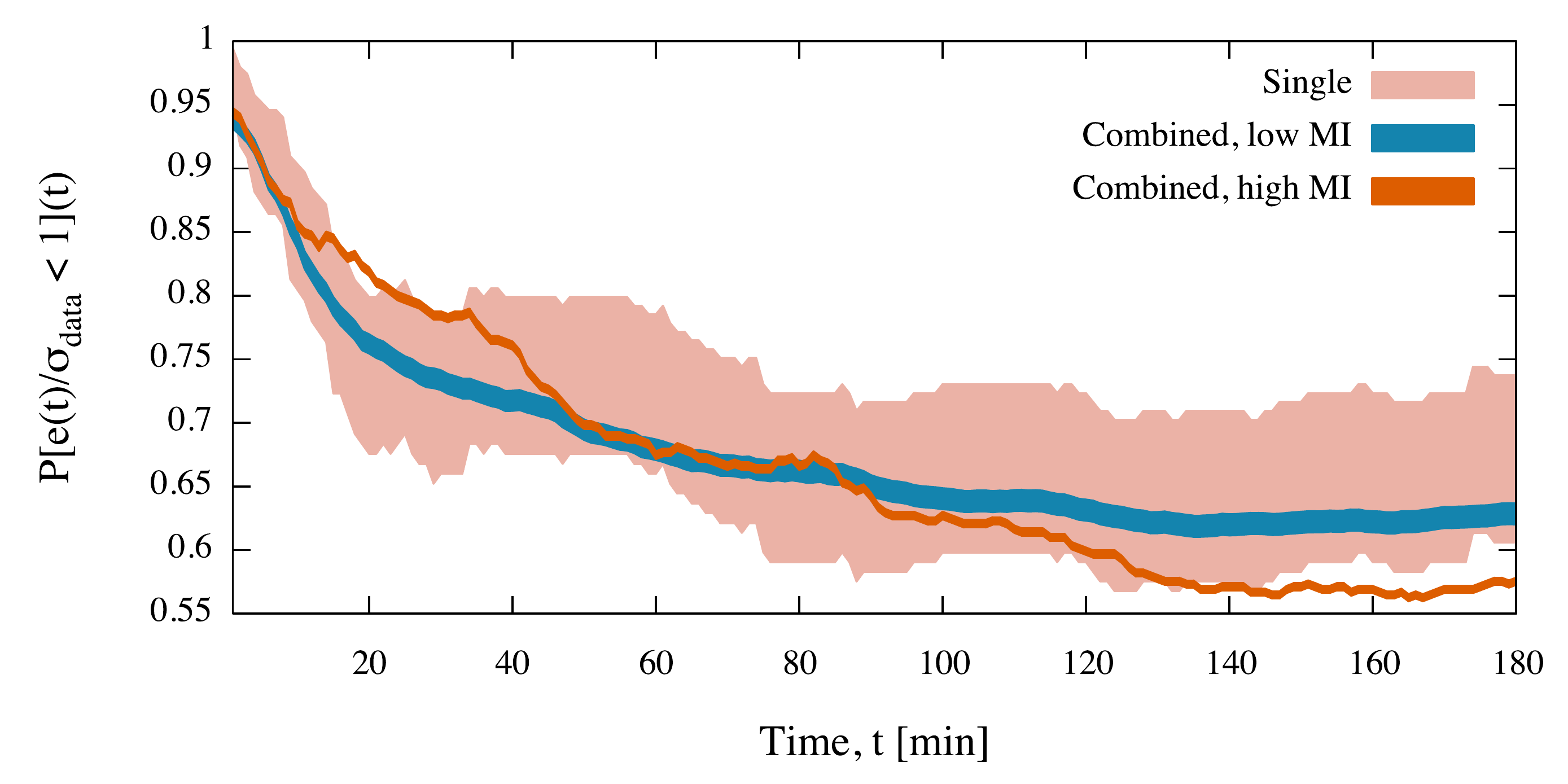}
\caption{\emph{Cabspotting dataset.} Temporal evolution of the fraction of mobility traces satisfying the condition $P(e(t) / \sigma_{data} < 1)$. Shaded areas indicate the 90\% confidence bands around the average value.}
\label{fig:prederr-cab-evol}
\end{figure}

Since the prediction is acceptable when the ratio $(e(t) / \sigma_{data})$ is below one, we investigate how the fraction of mobility traces satisfying this requirement, i.e., $P(e(t) / \sigma_{data} < 1)$, changes over time.
In Fig.\,\ref{fig:prederr-cab-evol} we show this temporal evolution, with 90\% confidence bands around the average values. Due to the lack of statistics (137 mobility traces) in the ``Single'' traces prediction, the bands are wider than for other cases (9316 mobility traces). All curves show decreasing behaviour for increasing prediction interval, as expected. In fact, the ``Combined, High MI'' predictor performs equal or better than others up to about 90 minutes. It is worth mentioning that the forecasting of every predictor is within the statistical uncertainity ($e(t) \leq \sigma_{data}$) for more than 50\% of mobility traces considered up to 3 hours.

\section{Discussion}
\label{sec:discussion}
In the context of mobile applications, the prediction of mobility patterns of
users is of great interest for several reasons. For instance, mobility
forecasting could be used to determine where the person will be and who he/she
will meet. Such an information can enable location-based mobile applications to
provide personalised services relating to the context the user is in.

% In addition, mobile devices could
%incorporate adaptive behaviours, determined by forecasted user activity (e.g.,
%saving battery when the user is expected to be away from home for a long time or
%trigger actions on other systems, such as home controllers). For these
%scenarios a high precision, both in the spatial and temporal domains, is of
%fundamental importance.

However, we are aware that there are scalability issues related to the
implementation and the deployment of the proposed technique. In particular, it
is well known that calculating mutual information in a multidimensional
environment (in this case, for a number of users larger than two) is
computationally expensive and does not scale efficiently. In fact, in this case
the computational complexity scales as $\mathcal{O}(N^n)$, where $N$ is the
subset of users and $n$ is the cardinality of the tuple taken into account.
However, the problem we are dealing with usually involves no more than 100
mobility traces (e.g., the size of the circle of most significant friends for
an individual). For this reason, we can still evaluate mutual information
values for any pair of traces, which scales as $\mathcal{O}(N^2)$. Nonetheless,
the multivariate embedding reconstruction is not feasible for a phase-space
larger than 40-dimensional. Even for a 2-coordinate signal representing a
mobility trace, it is not unusual to have a large embedding reconstruction due
to noisy data. Hence, no more than three mobility traces should be considered
simultaneously. 
Moreover, we are aware that the algorithm for searching the most suitable
additional set of mobility traces for improving the prediction scales as $\mathcal{O}(N^2)$,
where $N$ is the number of users taken into consideration in the application.

It is worth noting that many factors could be considered as signals of social
ties, according to the context of the deployment scenarios. As a consequence,
the quality of predictions might be deeply affected, either positively or
negatively, by the criteria used to detect social ties. In the Nokia MDC
dataset, we had no information about social ties between individuals, neither
of real nor virtual nature. In the available dataset, the presence of an
individual in the address book of another one actually represents the strongest
definition of a social tie. Moreover, two individuals with no social ties might
show similar mobility patterns, resulting in a high value of mutual
information. It is likely that individuals with strong social
ties (students, friends, co-workers and so on) behave similarly and their
mobility traces are characterised by patterns with a high value of mutual
information. Hence, the accuracy of the predictor will be improved in the case
the dynamics of traces is highly correlated, even if a social tie does not
exist.

A possible refinement of this work is the use of multivariate nonlinear
prediction with non-uniform embedding (different delays) and local polynomial
fitting~\cite{su2010prediction} in order to increase the accuracy of the
prediction.

\section{Conclusions}
\label{sec:conclusions}

In this paper, we have shown discussed multivariate nonlinear time series techniques can be successfully applied to improve the prediction of movement of users, by considering the movement of people with correlated mobility patterns.
More specifically, through the analysis of the Nokia Mobile Data Challenge traces, we have shown
that it is possible to exploit the correlation of social
interactions and user movement in order to improve the accuracy of forecasting of the future
geographic position of a user. By means of the Cabspotting dataset we have also shown
that when information about social ties is not available, mutual information can be
used to select pairs of users in order to improve prediction accuracy.

In other words, mobility correlation, measured by means of mutual information and
the presence of social ties can be used to improve movement forecasting by
exploiting mobility data of other individuals. This correlation can
be used as an indicator of potential existence of physical or distant social
interactions and vice versa.

%% The Appendices part is started with the command \appendix;
%% appendix sections are then done as normal sections
%% \appendix

%% \section{}
%% \label{}

%% References
%%
%% Following citation commands can be used in the body text:
%% Usage of \cite is as follows:
%%   \cite{key}          ==>>  [#]
%%   \cite[chap. 2]{key} ==>>  [#, chap. 2]
%%   \citet{key}         ==>>  Author [#]

%% References with bibTeX database:

\bibliographystyle{model1-num-names}
\bibliography{biblio}

%% Authors are advised to submit their bibtex database files. They are
%% requested to list a bibtex style file in the manuscript if they do
%% not want to use model1-num-names.bst.

%% References without bibTeX database:

% \begin{thebibliography}{00}

%% \bibitem must have the following form:
%%   \bibitem{key}...
%%

% \bibitem{}

% \end{thebibliography}

\end{document}